\documentclass[%
 reprint,
superscriptaddress,
 amsmath,amssymb,
 aps,
]{revtex4-2}
\usepackage{physics}
\usepackage{bbold}
\usepackage{amsmath}
\usepackage{latexsym}
\usepackage{graphicx}
\usepackage{amsthm}
\usepackage{hyperref}
\usepackage{lipsum}
\usepackage[english]{babel}
\usepackage{slashed}
\usepackage[compat=1.1.0]{tikz-feynman}
\usepackage{appendix}
\usepackage{subfig}
\usepackage{multirow}
\usepackage{mathtools}
\usepackage{subfig}
\usepackage{pgfplots}
\pgfplotsset{compat=1.15}
\usepackage{mathrsfs}
\usetikzlibrary{arrows}
\definecolor{ffffff}{rgb}{1,1,1}
\definecolor{qqqqff}{rgb}{0,0,1}
\definecolor{qqwuqq}{rgb}{0,0.39215686274509803,0}
\definecolor{zzttqq}{rgb}{0.6,0.2,0}
\definecolor{ttqqqq}{rgb}{0.2,0,0}
\definecolor{qqttqq}{rgb}{0,0.2,0}
\definecolor{qqccqq}{rgb}{0,0.8,0}
\usepackage{tikz-cd}
\usepackage{amsmath}
\usepackage{amsbsy}

\theoremstyle{plain}

\tikzset{
pattern size/.store in=\mcSize, 
pattern size = 5pt,
pattern thickness/.store in=\mcThickness, 
pattern thickness = 0.3pt,
pattern radius/.store in=\mcRadius, 
pattern radius = 1pt}
\makeatletter
\pgfutil@ifundefined{pgf@pattern@name@_b8864pxwu}{
\pgfdeclarepatternformonly[\mcThickness,\mcSize]{_b8864pxwu}
{\pgfqpoint{0pt}{0pt}}
{\pgfpoint{\mcSize+\mcThickness}{\mcSize+\mcThickness}}
{\pgfpoint{\mcSize}{\mcSize}}
{
\pgfsetcolor{\tikz@pattern@color}
\pgfsetlinewidth{\mcThickness}
\pgfpathmoveto{\pgfqpoint{0pt}{0pt}}
\pgfpathlineto{\pgfpoint{\mcSize+\mcThickness}{\mcSize+\mcThickness}}
\pgfusepath{stroke}
}}
\makeatother

\tikzset{
pattern size/.store in=\mcSize, 
pattern size = 5pt,
pattern thickness/.store in=\mcThickness, 
pattern thickness = 0.3pt,
pattern radius/.store in=\mcRadius, 
pattern radius = 1pt}
\makeatletter
\pgfutil@ifundefined{pgf@pattern@name@_f6j18z8qd}{
\pgfdeclarepatternformonly[\mcThickness,\mcSize]{_f6j18z8qd}
{\pgfqpoint{0pt}{0pt}}
{\pgfpoint{\mcSize+\mcThickness}{\mcSize+\mcThickness}}
{\pgfpoint{\mcSize}{\mcSize}}
{
\pgfsetcolor{\tikz@pattern@color}
\pgfsetlinewidth{\mcThickness}
\pgfpathmoveto{\pgfqpoint{0pt}{0pt}}
\pgfpathlineto{\pgfpoint{\mcSize+\mcThickness}{\mcSize+\mcThickness}}
\pgfusepath{stroke}
}}
\makeatother
\tikzset{every picture/.style={line width=0.75pt}} 

\begin{document}

\title{Do small massive superpositions necessarily significantly entangle with gravity?}
\author{Adrian Kent}
\affiliation{Centre for Quantum Information and Foundations, DAMTP, Centre for Mathematical Sciences, University of Cambridge, Wilberforce Road, Cambridge CB3 0WA, UK}
\affiliation{Perimeter Institute for Theoretical
	Physics, 31 Caroline Street North, Waterloo, ON N2L 2Y5, Canada.}
\date{\today}

\begin{abstract}
Christodoulou and Rovelli (CR) \cite{Christodoulou2019} have argued that a Bose et al.-Marletto-Vedral (BMV) experiment that confirmed the quantum nature
of gravity would give laboratory evidence for a quantum superposition of spacetime geometries created in the course of 
the experiment.  
Hanif et al. \cite{hanif2023testing} have argued that mass interferometers can be used to test whether gravity acts as a quantum entity when measured. 
We note that not all quantum models of gravity imply that mass superpositions necessarily become significantly entangled
with any degrees of freedom of the gravitational field during the experiments discussed.  
\end{abstract}
\maketitle

\section{Introduction}

Proposals by Bose et al. \cite{Bose2017} and Marletto-Vedral \cite{Marletto2017} (BMV) for experiments that could 
entangle two mesoscopic particles gravitationally have attracted a great deal of interest and discussion.
One key question is whether the experiments would give conclusive evidence for (or against) the quantum
nature of gravity.   Debate continues over whether 
there could be alternative explanations for the generation of entanglement (e.g.\cite{HR18,MV20,H21,HPF19,HVNCRI21,CMT21,martin2023gravity}).
It has also been noted that, in the original proposal in which entanglement witnesses are measured,
an apparent confirmation of entanglement could alternatively be explained by some local hidden variable
theory associated with gravity, although this specific loophole could be closed by 
extended versions of the experiment \cite{Kent2021a}.   

It has also been questioned whether the predictions of a full quantum gravity theory 
are necessarily well approximated by the Schr\"odinger equation with Newtonian potential
in these experiments, and hence whether we should in fact expect entanglement to be generated in any case \cite{privcomm}. 
We do not address this here: we adopt the consensus view in the literature and assume, for the following
discussion, that the approximation does make approximately the correct predictions for the matter degrees of freedom, and in particular that the particles do become strongly entangled for the experimental parameters proposed. 
If so, then the experiments are very strongly motivated whether or not 
they could definitively confirm quantum gravity, since they would certainly distinguish it from some
interesting alternatives, including gravitationally-induced collapse models \cite{K66,D84,D87,D89,P96,P98,P14,HPF19} and related ideas \cite{oppenheim2023postquantum} and
the possibility that a semi-classical gravity theory holds in the relevant regime \cite{Kent2021b}.   

Christodoulou and Rovelli \cite{Christodoulou2019} have argued that 
the predicted effect would evidence the quantum superposition of space-time geometries.
Their argument is based on a model that uses the Schr\"odinger equation and Newtonian potential, and so 
agrees with the BMV prediction that entanglement will be generated between the particles from an initially
separable state.   
Using a related model, Hanif et al. \cite{hanif2023testing} have recently
argued that mass interferometers can be used to test whether gravity acts as a quantum entity when measured. 

CR's discussion uses a model in which (1) degrees of freedom in the gravitational field become significantly entangled with a mesoscopic particle
in a mass interferometer, and (2) this entanglement essentially vanishes by the time the interfering paths recombine.   
The model is not a fully relativistic quantum theory 
of gravity and, at first sight, the combination of (1) 
and (2) appears hard to reconcile with relativistic
causality.   There are subtleties here, which we hope to
pursue elsewhere.   
The purpose of this note is to point out that (1) and (2) 
need not necessarily be correct.
Even given an analysis that  
uses the Schr\"odinger equation and Newtonian potential (or a causal 
extension thereof) and thus predicts
the generation of entanglement between the particles in a BMV experiment, there may be negligible entanglement between
the matter and gravitational degrees of freedom throughout the relevant experiments.

Our generalisation of CR's model, like the original, is open to criticism.  At best it outlines an intuition about the relevant underlying physics, not a proof of how it must behave.   
However, if it qualitatively captures the relevant physics, the degree to which a 
massive superposition becomes entangled with the gravitational field depends on 
pre-existing uncertainty in the gravitational field state.   
If so, while entanglement between matter and gravitational degrees of freedom can be created for superpositions of sufficiently distinct mass states, it need not
necessarily arise for superpositions of a sufficiently small mass in locations whose separation is sufficiently small, including perhaps any foreseeable mass interferometry. 
The relevant scales determining what is ``sufficiently distinct" and
``sufficiently small" should follow in principle from a full quantum gravity theory, including a theory of the initial conditions: in practice, they could be constrained empirically.   

\section{Do BMV experiments necessarily \emph{ create} quantum superpositions of geometries?}

Christodoulou and Rovelli (CR) \cite{Christodoulou2019} analyse the proposals by Bose et al. \cite{Bose2017} and Marletto-Vedral \cite{Marletto2017} (BMV) to 
entangle two mesoscopic particles gravitationally.  
CR argue that, in ``a generally covariant description of [BMV experiments], the relevant
effect turns out to be a quantum superposition of proper times" and ``$\ldots$ if General Relativity is assumed
to hold for masses at this [BMV] scale, measurement of this effect would count as evidence for
{\it quantum superposition of spacetime geometries}". 

They first note that if two mass $m$ particles are brought to a small separation $d$ for time $t$, each particle
acquires phase $\exp( - i E t/ \hbar)$, where the contribution for each particle of the other's gravitational
potential is $\delta E = G m^2 / d$, implying a phase shift $\delta \phi = ( G m^2 t )/ ( \hbar d ) $. 
This BMV effect is independent of $c$, and hence is a signature of a non-relativistic ($c \rightarrow \infty$) limit 
of quantum gravity.   Prima facie, this is consistent with using the non-relativistic 
Schr\"odinger equation and Newtonian potential, from which the phase shift is calculated.  
However, precisely what we should assume about quantum gravity models in this limit needs careful discussion.
CR aim to show that, ``on the assumption that General Relativity continues to hold for masses at the mesoscopic particle scale, detection of the BMV effect would provide evidence in favour of gravity being quantised in the sense that spacetime geometry obeys the superposition principle''.   A key issue here is whether an interpretation of the BMV effect’s
implications in the non-relativistic limit carries over to the relativistic regime.

CR's model supposes that in general the gravitational field has a quantum state 
belonging to a Hilbert space that contains what they call semiclassical states $\ket{g}$ that approximate classical geometries $g$, 
and also linear superpositions of such states.  It is worth stressing that, by semiclassical states, CR mean quantum states approximately describing
some classical geometry. They do not assume a semiclassical dynamical theory of gravity, in which the gravitational
field is determined by the expectation value of the quantum matter stress-energy.

CR further assume that before the BMV experiment the two particles and the gravitational field can be described by 
a product state of the form
\begin{equation}\label{product}
 \ket{\Psi_0} = \ket{\psi_1} \otimes \ket{\psi_2} \otimes \ket{g} \, .   
\end{equation}

In their description, the BMV experiment begins 
by splitting the state of each particle into the superposition of two semiclassical quantum states
\begin{equation}
 \ket{\psi_i} = {1 \over \sqrt{2}} ( \ket{\psi_i^L} + \ket{\psi_i^R} ) \, ,    
\end{equation}
which are distinguished by internal degrees of freedom (for example, spin) as well as location. 
Again, here, "semiclassical" means that the distribution approximates a classical matter distribution, in
this case that given by particle $i$ following the $L$ or $R$ path through the $i$th matter interferometer.

They then ``take the unrealistic simplification that the separation can be done very fast, say much faster than the time $d/c$'', 
where $d$ is the separation between the (nearest paths of) the two interferometers.   
Arguing that ``[i]mmediately after the split, the metric does not yet have time to change significantly'' they 
model the resulting state as 
\begin{eqnarray}\label{split}
\ket{\Psi_1} &=& {1 \over 2}(( \ket{\psi_1^L} + \ket{\psi_1^R} ) \otimes ( \ket{\psi_2^L} + \ket{\psi_2^R} ) ) \otimes \ket{g} \nonumber\\
&=& {1 \over 2}(\ket{LL} + \ket{RR} + \ket{LR} + \ket{RL}) \otimes \ket{g}  \, 
\end{eqnarray}
simplifying the notation by writing $\ket{XY} = \ket{\psi_1^X} \otimes \ket{\psi_2^Y}$.

They further argue that``[i]n a time of order $d/c$ the displacement of the particle produces a disturbance in the gravitational field that propagates at the speed of light to the distance of order $d$ (and past it) modifying accordingly.''
Taking the equations literally, this seems in tension with the assumption that the metric does not immediately change significantly: if gravitational field disturbances propagate at light speed then we expect the metric near the particle paths  {\it should} be affected by the time the split has taken
place.   Moreover, unless it is somehow suppressed or reversed, we should affect this 
disturbance to continue propagating at light speed away from the
experiment forever.   The matter components of (\ref{split}) correspond to different matter configurations with at least slightly
different gravitational fields, and the semiclassical fields $g$ are defined precisely. So, the statement that particle displacements produce gravitational field disturbances
also suggests that after the split there
should no longer be a single semiclassical metric, as suggested in (\ref{split}), but rather an entangled
superposition of particle and metric states. 

Another issue is that in a real experiment 
the metric $\ket{g}$ in (\ref{split}), or the metrics in any superposition, will depend (inter alia)
on quantum matter states throughout
and beyond the laboratory.  
There is also an apparent 
tension between the statement that gravitational disturbances propagate at light speed and describing the physics via a
non-relativistic model using Newtonian potentials, which propagate gravitational field effects instantaneously.

My understanding \cite{carlopriv} is that the model is intended to capture the correct physics qualitatively, while suppressing some physically significant features that would be evident in a complete analysis.  In other words, the model is not meant to be taken literally. 
These apparent tensions thus deserve more careful 
discussion than we give here.  Our aim here
is neither to 
justify nor refute CR's qualitative model but to note generalisations 
that have qualitatively different features.  

Although CR (taken literally) suggest that the state immediately after the split is (\ref{split}), they
agree that their model implies that the matter and metric states become entangled during the experiment:
indeed, this is a key point in their argument.
They write $\ket{g_d}$ as shorthand for the metric corresponding to the two particles separated 
by distance $d$.   This notation implicitly assumes that the four separations in the experiment are distinct and
suppresses dependence on other position degrees of freedom.   In the real world, given an approximately fixed background
semiclassical spacetime, the particle locations as well as their separation matter, of course: we can distinguish the state in which
the two particles are at respective positions $\underline{x}, \underline{x}+\underline{d}$ from that in which they are at $\underline{x'}, \underline{x'}+\underline{d}$, where $| \underline{d} | =d $, by
referring to the background gravitational fields of Earth, other celestial bodies, laboratory apparatus and so on. 
We discuss later whether we {\it should} assume a single approximately fixed background semiclassical spacetime, but
note that in any case we can identify the particles' positions relative to the Earth and laboratory with high
precision, so a full description of the physical state during the experiment cannot be invariant under 
a single translation of both particles, as Eqns. (\ref{split},\ref{during}) would suggest if taken literally.  

We continue to follow CR's implicit assumptions and notation in the following, writing the separations
in the different branches as $d_{LL}, d_{LR}, \ldots$ with respective associated metrics $\ket{g_{d_{LL}}}, \ket{g_{d_{LR}}}, \ldots$. 
CR then argue that the state during a BMV experiment evolves to 
\begin{eqnarray}\label{during}
\ket{\Psi_2}&  = 
{1 \over 2}(\ket{LL} \ket{g_{d_{LL}}} + \ket{RR} \ket{g_{d_{RR}}} + \\
&\qquad \ket{LR} \ket{g_{d_{LR}}} +\ket{RL} \ket{g_{d_{RL}}} ) 
\, \nonumber . 
\end{eqnarray}
It will be convenient for our discussion to fix a time, $t_2$, in lab frame
at which CR suppose this state to arise.   
As they note, ``[i]n this state the metric is not semiclassical anymore. It is in a superposition of macroscopically distinct semiclassical states, entangled with both particles.''
Indeed if, as CR argue, the spacetimes are distinguishable with certainty, the states $\ket{g_{d_{XY}}}$ for different $XY$ should be orthogonal, and the gravitational field state is thus maximally entangled with the matter state.   
Note nonetheless though that Bose et al.’s \cite{Bose2017} perturbative quantum gravity analysis suggests that there is almost zero entanglement between the matter and radiative gravitational degrees of freedom throughout the experiment.   

CR next make the simplifying assumption that $d_{RL}=d$ is much smaller than the other three separations, which
are too large for their respective Newtonian potentials to create any significant phase shift during the experiment.
They derive the phase difference between the RL branch and the others, 
\begin{equation}
\delta \phi = {{ G m^2 t} \over {\hbar d }} \, , 
\end{equation}
in agreement with the BMV expression \cite{Bose2017,Marletto2017}, 
by calculating the gravitational redshift of the proper time along the particles' worldline. 
In their model, this produces the state
\begin{eqnarray}\label{aftert}
\ket{\Psi_3}  = &
{1 \over 2}(\ket{LL} \ket{g_{d_{LL}}} & + \ket{RR} \ket{g_{d_{RR}}} + \ket{LR} \ket{g_{d_{LR}}} \\
& &  + \exp (i G m^2 t / \hbar d ) \ket{RL} \ket{g_{d_{RL}}} ) 
\, \nonumber . 
\end{eqnarray}
This takes place at time $t$; for our discussion we write $t=t_3$.
This is the basis for CR's claim that a BMV experiment creates {\it de novo} a superposition of spacetimes.   

When the two components of each particle are recombined, CR argue that the metric evolves back to the same state $g$
in each branch, and so model the combined state as
\begin{equation}\label{end}
\ket{\Psi_4}  = {1 \over 2}(\ket{LL} + \ket{RR} + \ket{LR}  + \exp (i G m^2 t / \hbar d ) \ket{RL} )\ket{g}  
\, . 
\end{equation}
If the experiment had duration $t=\pi \hbar d/ G m^2$ this gives
\begin{equation}\label{endsimple}
\ket{\Psi_4}  = {1 \over 2}(\ket{LL} + \ket{RR} + \ket{LR}  - \ket{RL} )\ket{g}  
\, . 
\end{equation}
We write $t_4$ for the time at which this is supposed to occur in CR's model.
At the level of discussion given in CR, the argument is that the recombined components have the same Newtonian potential and ``hence'' the 
same metric state $g$: this is discussed in more detail in \cite{chen2023quantum}.    

A concern is that this result is an artefact of using the non-relativistic Newtonian
potential in a model that is intended to represent the quantum states of relativistic space-times.    
Since the Newtonian potential alters instantaneously everywhere, recombining the particles causes it
to ``forget'' that they were separated.  As we noted earlier, this is in tension with CR's 
statement that particle displacements cause disturbances in the gravitational field that propagate
at light speed.   If we incorporated this into the model, each of the four pairs of paths would have
a different associated metric even after recombination, since the effects of the displacements while
separated continue to propagate indefinitely at light speed in the metric.  

To see this more explicitly, consider a crude alternative model.    Suppose that a slow-moving (i.e. non-relativistic)
particle has trajectory $x(t)$ in the laboratory rest frame.   We take the associated potential at point $y$
and time $t'$ to be 
\begin{equation}\label{causalp}
\Phi (y, t') = { - { G m} \over { | y - x(t_0 ) | }} \, ,
\end{equation}
where $t_0$ is determined by the constraint $t_0 < t'$ and 
\begin{equation}
   { { | y - x(t_0 |} \over { |t' - t_0 |}} = c  \, . 
\end{equation}
A semiclassical (in CR's sense) metric state $\ket{g}$ determined by this potential is affected by the path
separation at all later times, even after path recombination.

This would imply that the four paths in the BMV experiment produce distinct metric states, which remain
distinct and orthogonal through to the end of the experiment, so that even as the matter paths are recombined the joint state has a similar form to (\ref{aftert}),
with the different metric states reflecting the different path histories.   

If this {\it were} qualitatively correct, since matter and gravitational states remain entangled, 
BMV entanglement tests on the matter states would fail.
A similar conclusion would hold if we consider only the two paths in a {\it single} matter interferometry experiment; matter
interferometry would fail to show interference.
This illustrates again that taking CR's qualitative model too literally leads to contradictions.   

We plan to look in more detail at the physics underlying CR's model elsewhere.  
Here we simply wish to note one interesting possibility that is relevant to these issues: that CR's assumption that the metric state is initially semi-classical is significantly incorrect.
By this we mean not just the obvious statement that it is not precisely true, but 
that it is significantly quantitatively incorrect in a way that leads to qualitatively incorrect conclusions in both their model and the alternative with potential (\ref{causalp}).

Now, it should be uncontroversial that the metric state cannot be {\it precisely}  semiclassical because the
gravitational field is generated by all the masses in the past light cone, each of which has some position spread, and should also have its own independent spread arising from the initial conditions and subsequent fluctuations.   
We can refine CR's model, while retaining their assumption that semiclassical metric states form a basis, to allow for this, replacing their equation (\ref{product}) by
\begin{equation}\label{smeared}
    \ket{\Psi'_0 }= \ket{\psi_1} \otimes \ket{\psi_2} \otimes \int \rho(g) \ket{g} d \mu (g) \, .   
\end{equation}
Here $\rho(g)$ is a wave function and $\mu(g)$ a measure on spacetimes, 
with 
\begin{equation}
\braket{g}{g'} = \delta (g , g' ) \, \quad \int \delta(g,g') d \mu (g) f(g)  = f (g') 
\end{equation}
and 
\begin{equation}
\int | \rho (g) |^2 d \mu (g) = 1 \, .
\end{equation}
We suppose that $| \rho (g) |^2 $ has a significant spread that would
remain roughly constant over the timescale of the experiment, in the 
absence of any interferometry.  

In principle, if this model were justifiable in the relevant regime, $\rho(g)$ would
be derived from the initial quantum state of the universe and 
its evolution law, and $\mu(g)$ from a complete quantum theory of gravity.   
For this discussion we treat $\rho$ and $\mu$ as unknown, though $\mu$ is constrained by symmetries and other theoretical principles and
some bounds on the spread of $\rho$ could be estimated (given assumptions) by decoherence models.  Absent these discussions, some of the relevant 
properties of $\rho$ and $\mu$
could be inferred empirically (again, in any regimes where the model is justifiable).   

Linearity of evolution then gives us at time $t_2$
\begin{eqnarray}\label{duringp}
\ket{\Psi'_2}&  = 
{1 \over 2}(\ket{LL} \int \rho^{LL}_2 (g) \ket{g} d \mu (g) + \\
&\qquad  \ket{RR}  \int \rho^{RR}_2 (g) \ket{g} d \mu (g) + \nonumber \\
&\qquad  \ket{LR} \int \rho^{LR}_2 (g) \ket{g} d \mu (g) + \nonumber \\
& \qquad \ket{RL}  \int \rho^{RL}_2 (g) \ket{g} d \mu (g)    ) 
\, \nonumber 
\end{eqnarray}
in place of (\ref{during}).
Here $\rho^{XY}_2 (g)$ is the wave function at time $t_2$ for a gravitational field initially in superposition 
\begin{equation}\label{gravinit}
\int \rho(g) \ket{g} d \mu (g)    
\end{equation}
when the particles follow trajectory $XY$.
Whereas (\ref{during}) is a superposition of four distinct matter and spacetime states, 
\ref{duringp} is a continuous superposition.
The state (\ref{during}) is maximally entangled because
\begin{equation}
    \braket{g_{d_{XY}}}{g_{d_{X'Y'}}}  = \delta_{XX'} \delta_{YY'} \, .
\end{equation}
The degree to which (\ref{duringp}) is entangled depends on the quantities
\begin{equation}
    \int \bar{\rho}^{XY}_2 (g) \rho^{X'Y'}_2 (g) d \mu (g) \, , 
\end{equation}
which need not necessarily be small for $XY \neq X'Y'$.
In particular, if the spread of $\rho (g)$ -- which, recall, is a property of the initial state before the experiment -- is large compared to the gravitational contributions of the particles, they may be close to one.
Note that this can hold true both in CR's original model with the instantaneously propagating
Newtonian potential and in our alternative model with the potential (\ref{causalp}) that effectively propagates at light speed.

We now assume that the spread of $\rho (g)$ is indeed sufficiently large that 
\begin{equation}
  \int \bar{\rho}^{XY}_2 (g) \rho^{X'Y'}_2 (g) d \mu (g) \approx \int \bar{\rho}^{XY}_2 (g) \rho_2 (g) d \mu (g) \approx  
  1 \, 
  \end{equation}
for all  $XY, X'Y'$.
The intuition here is that $\rho_2 (g)$ is a wave function that would be obtained at 
time $t_2$ from some notional averaged state of the experiment, such as the particles following the
median of the two paths in each wing, with the initial spread in $\rho (g)$ 
dominating compared to the differential contributions between any pair of paths
and the median (or any pair and any other pair).

We then have 
\begin{equation}\label{duringps}
\ket{\Psi'_2}  \approx 
{1 \over 2}(\ket{LL} + \ket{RR}  + 
 \ket{LR} + \ket{RL} ) \int \rho_2 (g) \ket{g} d \mu (g)    ) 
\, , 
\end{equation}
Evolving further to times $t_3$ and $t_4$ we get similar expressions
\begin{eqnarray}\label{aftertp}
\ket{\Psi'_3}  \approx &
{1 \over 2}(\ket{LL} + \ket{RR}  + \ket{LR} 
 + \exp (i G m^2 t / \hbar d ) \ket{RL} ) \times \nonumber\\
 & \int \rho_3 (g) \ket{g} d \mu (g)   
\, , 
\end{eqnarray}
\begin{eqnarray} \label{endp}
\ket{\Psi'_4}  \approx & {1 \over 2}(\ket{LL} + \ket{RR} + \ket{LR}  + \exp (i G m^2 t / \hbar d ) \ket{RL} ) \times \nonumber\\
& \int \rho_4 (g) \ket{g} d \mu (g)  
\, . 
\end{eqnarray}
This last state has essentially no entanglement between the matter and gravity states, and so allows the 
usual prediction that matter entanglement should be generated and witnessable in a BMV experiment.
The same conclusion holds, redefining $\rho$ and the $\rho_i$ suitably, 
in our model with potential (\ref{causalp}). 

A similar analysis gives the usual prediction that matter interference should be 
observed in a single particle mass interferometer.  Again, this holds both
with the standard Newtonian potential and in our alternative model with
potential (\ref{causalp}). 

In all these cases the particle trajectories do affect the state of the
gravitational degrees of freedom, even after the end of the experiment, but the effect is (by hypothesis) negligible compared
to the gravitational state's initial spread.    

\section{Discussion}

Pace CR’s model, discrete superpositions of distinct semi-classical spacetimes need not necessarily be created {\it de novo}
in a BMV experiment.   There is an alternative, namely that the initial gravitational field state is already a (continuous) superposition
of distinct space-times.  
This alternative is also consistent with a crude semi-relativistic alternative version of CR's model that incorporates the light speed propagation of gravitational
effects. 

If this alternative is qualitatively correct, superpositions of distinct space-times do exist in nature, but need not necessarily be discrete superpositions of the form (\ref{aftert}), and need not necessarily be created {\it de novo} in the course of the experiment.

\bigskip

\section{Does measuring the nearby gravitational field necessarily decohere a mass interferometer?}

Hanif et al. \cite{hanif2023testing} consider a mass interferometry experiment in which the nearby gravitational
field either is or is not measured by a probe, which may be a second mass interferometer.    
It is crucial in their setup that the probe is in a region near to, but not overlapping with, the mass interferometer, 
and sufficiently separated that (to very good approximation) the only relevant interaction is between the probe
and its local gravitational field.  In particular, as in BMV experiments, the separation must be such that
the effect of any Casimir-Polder interaction is relatively negligible.   

Appealing (inter alia) to Christodolou-Rovelli's arguments, discussed above, they suggest that, if the 
gravitational field is quantum, its measurement by the probe should distinguish between the gravitational
field states associated with the massive particle following the two possible paths through the interferometer.
Effectively, the probe measurement determines which path
the massive particle follows through the interferometer.   
Hence, they argue, the mass interferometer will be decohered when the probe is present, while it should
show interference when the probe is absent.   
On the other hand, they note, in non-standard alternative theories in which the mass interferometer generates a 
single classical gravitational field in its neighbourhood, a probe measurement of this field will not alter
the state of the particle in the mass interferometer, and interference will be observed whether or not the 
probe is present.   For example, this would be the case if semi-classical gravity were to hold in the relevant
regime (i.e. for delocalized superpositions of massive particles with the mass and separations involved in the 
mass interferometer).   Similarly, it would be true in hybrid theories in which a classical gravitational field
is generated stochastically by quantum matter states in the relevant regime.    The key point, they argue, is 
that a classical gravitational field, by definition, can in principle be measured with arbitrarily precision
without disturbance, whereas measuring a quantum gravitational field necessarily disturbs it if the outcome
of the measurement was initially unpredictable.    
Hence their proposed experiment distinguishes between quantum theories of gravity and classical alternatives.

We question whether their experiment would necessarily cause a 
failure of interference if gravity is quantum.   If 
the gravitational field at the start of the interference is spread,
as in (\ref{smeared}), so that the initial state is 
\begin{equation}\label{smeared2}
    \ket{\Phi'_0 }= \ket{\phi_1}  \otimes \int \rho(g) \ket{g} d \mu (g) \, ,  
\end{equation}
then the state after splitting the interferometer beams, at time $t_2$, has the form
\begin{equation}\label{duringps2}
\ket{\Phi'_2}  \approx 
{1 \over 2}(\ket{L}  \int \rho^L_2 (g) \ket{g} d \mu (g)  +
\ket{R} \int \rho^R_2 (g) \ket{g} d \mu (g)) 
\, ,
\end{equation}
where $\rho^X_2 (g)$
is the wave function at time $t_2$ for a gravitational field initially in superposition 
\begin{equation}\label{gravinit2}
\int \rho(g) \ket{g} d \mu (g)    
\end{equation}
when the particle follows trajectory $X$.
The degree to which (\ref{duringps2}) is entangled depends on 
\begin{equation}\label{overlap}
    \int \bar{\rho}^{X}_2 (g) \rho^{Y}_2 (g) d \mu (g) \, , 
\end{equation}
 for $X \neq Y$, which need not necessarily be small.
In particular, if the spread of $\rho (g)$ is large compared to the gravitational contributions of the particles, the overlap (\ref{overlap}) may be close to one.
A measurement of the gravitational field, even if it collapses the gravitational
field to (in CR's sense) a semi-classical state $\ket{g}$, would then
produce essentially no decoherence of the interferometer.  
In this case the experiment would not distinguish between quantum theories
of gravity and the classical alternatives Hanif et al. have in mind.

Although it is tangential to the main point of our discussion, we also question 
the other point of their argument, that measuring classical gravitational
fields necessarily has no effect on the quantum state that effectively
generated them.   Measuring a classical gravitational field with a massive probe
necessarily alters that field in the region of the probe, since the probe
has its own gravitational field, and the measurement process involves some
matter displacement and generally a transfer of energy between the gravitational field and matter.   This influence propagates, presumably
at light speed in a relativistic theory, to have a nonzero effect eventually
on the gravitational field throughout the future light cone of the probe. 
In particular, it very swiftly affects the field in the region of the 
interferometer, since although the probe is separated from the interferometer
it is very close.  It seems at least logically possible that, in some
hybrid quantum-classical matter-gravity theories, the classical gravitational
field affects the quantum matter not just by influencing its evolution in the standard way, but more dramatically, 
so that measurement-like perturbations
of the gravitational field indirectly produce measurement-like collapses
of the quantum matter state.
That said, this last possibility appears at first sight conspiratorial, 
without evident independent theoretical motivation.    

In contrast, within the framework of the models under discussion,  it seems possible that the initial state might be
modelled by (\ref{duringps2}) with an 
(\ref{overlap}) might be large for $X  \neq Y$, i.e., for the two different paths.   It would be very interesting to develop
quantitative estimates within models of quantum gravity and
decoherence.

\section{Discussion}

Our main point is that Hanif et al.'s experiment need not necessarily give the outcome they predict if gravity is quantum.  In a quantum gravity theory whose relevant features are modelled
by (\ref{smeared2}-\ref{gravinit2}), with the expression (\ref{overlap}) large, 
measuring the gravitational field by a probe will not necessarily decohere
the interferometer significantly.  

If the experiment did in fact distinguish between the probed and unprobed cases, 
we would agree that this constitutes good evidence for the quantum nature of
the gravitational field, since the most obvious explanation is a quantum gravity
model for which (in the language of ours)  (\ref{overlap}) is small.
For the reasons noted above, we would not take the evidence as quite
conclusive, because alternative explanations are logically possible within hybrid models in 
which gravity is classical.   However, at present we see no clear theoretical motivation
for hybrid models with the specific feature required.  

\section{Conclusions}

Neither CR nor Hanif et al. assume a specific model of quantum gravity, such as string theory,
for their arguments, which are intended to identify consequences of any plausible quantum
gravity theory.    However they {\it do} implicitly assume properties of the initial
(pre-experiment) state of the gravitational field, which
may be qualitatively correct but need not necessarily be.  
Hanif et al.'s proposed experiment would give very valuable evidence on these questions.

\section{Note added}

Hanif et al. have revised and clarified their discussion \cite{hanif2024testing} in
response (inter alia) to an earlier draft of this note. 
I thank them for clarificatory comments.  

\bigskip

\section{Acknowledgements}
I acknowledge financial support from the
UK Quantum Communications Hub grant no. 
EP/T001011/1. 
This work was supported in part by Perimeter Institute for
Theoretical Physics. Research at Perimeter Institute is supported by
the Government of Canada through the Department of Innovation, Science
and Economic Development and by the Province
of Ontario through the Ministry of Research, Innovation and Science.
I am very grateful to Carlo Rovelli for many invaluable comments
and insights.   I also thank Samuel Fedida for helpful comments.

\bigskip

\bibliographystyle{unsrt}
\bibliography{library,postdocbiblioshort}

\end{document}